\documentclass[%
 reprint, 
nofootinbib,
 amsmath,amssymb,
 prl
]{revtex4-2}

\usepackage{graphicx}
\usepackage{dcolumn}
\usepackage{bm}
\usepackage[mathlines]{lineno}

\newcommand{\OO}{\mathcal{O}}

\newcommand{\ex}[1]{\langle #1\rangle}

\newcommand{\RR}{\mathbb{R}}

\newcommand{\maxprob}[2]{\begin{array}{ll}
	\operatorname{maximize} & #1 \\
	\text { subject to } & #2 \\
\end{array}}

\usepackage{xcolor}

\begin{document}


\title{One-dimensional reflection in the quantum mechanical bootstrap}

\author{David Berenstein}
 \email{dberens@physics.ucsb.edu}
 \affiliation{Department of Physics, UC Santa Barbara 93106}
\author{George Hulsey}%
 \email{hulsey@physics.ucsb.edu}
\affiliation{Department of Physics, UC Santa Barbara 93106}%




\date{\today}

\begin{abstract}
We describe the application of the quantum mechanical bootstrap to the solution of one-dimensional scattering problems. By fixing a boundary and modulating the Robin parameter of the boundary conditions we are able to extract the reflection coefficient for various potentials and compare to physical expectations. This includes an application of semidefinite programming to solving a half-line Schrodinger problem with arbitrary Robin boundary conditions. Finally, the WKB approximation is used to numerically determine the scattering behavior of the exponential potential of Liouville theory.
\end{abstract}

\maketitle



\section{Introduction}
\label{sec:intro}
The numerical bootstrap has been the subject of much recent work \cite{Lin:2020mme,Han:2020bkb,Aikawa:2021eai,Kazakov:2022xuh,kazakovMM,Lawrence:2021msm,PhysRevLett.108.200404}, applying techniques developed for the conformal bootstrap to other quantum-mechanical systems. In our previous work, we described the application of the numerical bootstrap to the solution of one-dimensional Hamiltonian bound-state problems with arbitrary wavefunction boundary conditions \cite{Berenstein:2021dyf,Berenstein:2021loy,Berenstein:2022ygg}. Here, we extend these ideas to the characterization of one-dimensional scattering. The problem of scattering in one dimension has been studied from multiple perspectives, from the development of approximation schema \cite{russian1,russian2,eikonalMoller,schiff1956approximation,saxon1957theory} to theorems relating the scattering phase shift to properties of the bound state spectrum \cite{boya2008quantum,kiers1996scattering,dong2000levinson}, though the focus of the literature on scattering is mainly on three dimensions.

For many applications of the quantum mechanical bootstrap to work, the moments of the probability distribution need to be finite, but the scattering problem does not have this property. The essential idea of this letter is that the scattering behavior of a potential may be determined by solving a related family of bound-state problems with the same potential, a family parametrized by the boundary conditions at some location. One places a geometric cutoff in the far scattering region and then solves the bound-state problem with all the possible boundary conditions at that cutoff. To determine the scattering phase, one uses the same boundary condition to match to the asymptotic (plane-wave) wavefunction. In this asymptotic region, if the potential is not constant, we supplement the plane-wave solution by a WKB approximation to the desired order of precision. 

We describe how one can determine scattering phase shifts using any numerical eigen-energy algorithm. In particular, these include the semidefinite programming (SDP) bootstrap algorithm of our previous work  now applied to problems on the real half line \cite{Berenstein:2022unr}. This is similar in spirit to the program of determining scattering data by the energy spectrum at  finite volume, which can be realized on a lattice computation \cite{Luscher:1986pf}.

To begin, consider a one-dimensional quantum system with a one-sided potential such that $V(x) = 0$ for $x \leq 0$ and $V\to \infty$ as $x \to \infty$ and with the Hamiltonian  given by
$$
H= -\partial_x^2+V(x).
$$ 
Physically, one understands that waves come in from $x = -\infty$, reflect off the potential barrier, and return to $x = -\infty$ with a  phase shift $\delta(E)$ depending on the incident energy $E$. In this one-dimensional, one-sided system, this phase shift completely specifies the $S$-matrix and hence the physics of scattering.

Let $R(E) = \exp(i\delta)$ be the reflection coefficient for waves with energy $E = k^2$. One has
\begin{equation}
    \label{eq:planewave}
    \psi(x)  = A\left[e^{ikx} + R(E)e^{-ikx}\right];\quad x \leq 0.
\end{equation}
Imagine placing a hard boundary at $x = 0$ and enforcing there a Robin boundary condition
\begin{equation}
    \label{eq:RobinBC}
    \psi(0) + a\psi'(0) = 0 \implies a = -\frac{\psi(0)}{\psi'(0)}.
\end{equation}
Here $a$, the Robin parameter, is real `plus infinity'-valued; the point $a = \infty$ corresponds to the Neumann condition. The variable $a$ parametrizes self-adjoint extensions of half-line Schrodinger operators \cite{reed2}. 

On the right hand side $x \geq 0$, the boundary condition \eqref{eq:RobinBC} defines a self-adjoint Hamiltonian and therefore a discrete spectrum of energies which changes continuously with $a$. Each value gives a spectrum $E_n(a)$ which may be inverted to give a function $a(E)$. One then matches the value of the wavefunction and its derivative at the boundary $x = 0$. This gives a direct relationship between the reflection coefficient $R(E)$ and the Robin parameter $a(E)$:
\begin{equation}
    R(E)=\frac{i ka(E)+1}{i ka(E)-1};\quad k = \sqrt{E} > 0.
\end{equation}
One can therefore extract the $S$-matrix by solving a succession of bound state problems for different values of the Robin parameter $a$.  Just as bound states can be considered as a suitably interfering set of plane waves, the data of plane wave scattering are equivalent to a continuous family of bound states.

In this paper we demonstrate this approach to determining the scattering phase shift of numerous potentials. We do so by applying the semidefinite programming bootstrap algorithm  as well other numerical methods to determine the energy spectrum for half-line problems with arbitrary Robin conditions, and from there to extract the expected reflection coefficients. We demonstrate the consistency of this approach by comparing the numerics to physical expectations for scattering. Finally, we use WKB methods to extend the approach to potentials which are not constant on the scattering region but that are instead slowly varying in the asymptotic regime and for which $V(x)<k^2$ in the $x<0$ region. We show  as an example that we can numerically reproduce the reflection coefficient of the zero-mode Liouville theory by this approach. 

\section{Half-line SDP problems}
Semidefinite programming (SDP) is a class of convex optimization problems where the feasible domain is a subset of the cone of convex matrices. Their application to the bootstrap program has long been recognized, and as a numerical tool it has been applied with great success in the conformal bootstrap, the lattice Yang-Mills bootstrap, and in other spin-chain lattice theories \cite{simons-duffin,Kazakov:2022xuh,Lawrence:2021msm}. 

In our previous work \cite{Berenstein:2022unr} we described how to formulate the one-dimensional quantum mechanical bootstrap as a semidefinite program. We showed how this could be applied to determine the spectrum of a polynomial potential of arbitrary degree on the real line. The generalization of this approach to half-line problems is straightforward, though one must include the anomaly terms discussed in \cite{Berenstein:2022ygg} to properly handle the boundary conditions. 

To see the role of the anomalies, consider a potential $V(x)$ on the real half-line $x \geq 0$ and with a general Robin boundary condition at the origin $x =0$, so that $\psi_0 + a\psi'_0 = 0$, writing $\psi_x \equiv \psi(x)$. 
To set up the semidefinite program one must first compute (recursively) some positional moments $\ex{x^k}_\psi$ in some undetermined state $\psi$. Relations between these moments are furnished by the bootstrap constraints. Let $H$ be the Hamiltonian and $\OO$ some operator; these constraints, true in energy eigenstates, are of the form
\begin{equation}
    \label{eq:basicBtspconstraint}
    \ex{[H,\OO]} + \ex{(H^\dagger - H)\OO} = 0,
\end{equation}
where the second term is the anomaly. The anomaly vanishes if $\OO$ leaves the domain of $H$ invariant; these issues are discussed in depth in \cite{Berenstein:2022ygg,halflinemom,anomaly,anomaly1}. 

Many of these anomalies do happen to vanish. If the wavefunction associated with the eigenstate is $\psi_x$, the only nonzero anomalies\footnote{If the boundary condition were at some $x \neq 0$, the anomalies would continue to persist at higher orders in the recursion.} are given by
\begin{gather}
    \mathcal{A}[nx^{n-1}] = -\frac{1}{M}\psi_0^2\delta_{n,1},\\
    2 i \mathcal{A}[xp]=-\frac{1}{M} \psi_0 \psi_0^{\prime},\\
    2 i \mathcal{A}[p]=\frac{1}{M}\left(\psi_0^{\prime}\right)^2+2 \psi_0^2(E-V(0)).
\end{gather}
The last of the anomalies appears in the constraint \eqref{eq:basicBtspconstraint} for the operator $\OO  = p$ as:
\begin{equation}
    0 = -\ex{V'(x)} + 2i\mathcal{A}[p].
\end{equation}
This is a quantum-corrected version of Newton's second law which now includes an anomalous contribution from the boundary condition. With a polynomial potential $V = \sum^d c_m x^m$ and Robin boundary conditions at $x = 0$, this lowest-level constraint allows one to determine the boundary value $\psi_0^2$ in terms of the other moments:
\begin{equation}
    \label{eq:boundaryvalue}
    \psi_0^2 = \left[2(E-c_0) + \frac{1}{a^2 M}\right]^{-1}\cdot \sum_{m = 1}^d 2mc_m \ex{x^{m-1}}.
\end{equation}
where we have additionally used the relation $\psi_0' = -\psi_0/a$. 

One can therefore express all the nonzero anomalies as linear combinations of moments $\ex{x^k}$ of order $k \leq d-1$. Higher moments $\ex{x^k}$ with $k \geq d$ can then be computed recursively using the constraints \eqref{eq:basicBtspconstraint}. Like in the real line case, the higher moments may be written as linear functions of the $d-1$ primal moments $\ex{x},\ex{x^2},\ldots,\ex{x^{d-1}}$. Their coefficients will depend (non-linearly) on the energy $E$ and the Robin parameter $a$. 

The formulation and solution of the SDP then proceeds as described in \cite{Berenstein:2022unr}, with the salient modification being the required positivity of two $K \times K$ moment matrices:
\begin{equation}
    M_{ij} = x_{i+j},\ M'_{ij} = x_{1+i+j};\quad 0 \leq i,j,\leq K-1.
\end{equation}
One maximizes the minimal eigenvalue of both of these matrices simultaneously, defining the optimization problem
\begin{equation}
\label{eq:halfSDP}
    \maxprob{t}{\left[\begin{array}{cc}
M(x_i) & 0 \\
0 & M'(x_i)
\end{array}\right] - t \operatorname{Id} \succeq 0,}
\end{equation}
at fixed energy $E$ and Robin parameter $a$. This is a linear semidefinite program with $d$ primal variables $(t,x_1,\ldots,x_{d-1})$ written in the dual or linear-matrix-inequality form. Pairs $(E,a)$ are deemed physically allowable at size $K$ if the optimal objective $t_\star$ is positive.
The convergence property of the bootstrap ensures that if a pair is unphysical at some $K$ it will continue to be unphysical at all higher matrix sizes.


It should be noted that the \textit{Mathematica} function \verb|NDEigensystem| now supports Robin (or generalized Neumann) boundary values. At the level of machine precision, this finite element-based algorithm runs orders of magnitude more efficiently than the SDP algorithm described here, which runs on a multiple-precision solver \cite{sdpa}. As a demonstration of the application to half line problems, some of the data in the following sections is generated via the bootstrap algorithm. However, the discussion of the scattering phenomenology and role of the Robin parameter is independent of the algorithm used. It depends only on being able to numerically solve the spectrum of a one-sided differential eigensystem with a Robin boundary condition. 

\section{Bootstrapping pure reflection}
In the case of pure reflection, the quantity of interest is the reflection coefficient. In terms of the wavenumber $k = \sqrt{E} > 0$ and Robin parameter $a$ it is given by
\begin{equation}
    R(k)=\frac{ika +1}{ika-1} = e^{i\delta(k)}.
\end{equation}
The phase shift is the complex argument of $R(k)$; using the identity
\begin{equation}
    \label{eq:identity}
    \arg \frac{1+it}{1-it} = 2 \tan^{-1}(t),
\end{equation}
one may write the phase shift $\delta(k)$ as
\begin{equation}
    \label{eq:generalPhase}
    \delta(k) = 2\tan^{-1}\left(ka\right).
\end{equation}
By determining $a(E)$ one has determined the phase shift, and hence all the information about scattering off the one-dimensional, one sided potential. 
\subsection{The half-harmonic wall}
In the case of a half-harmonic potential wall one can compute the function $a(E)$---and therefore the phase shift---exactly; we compare this to data obtained by solving a succession of bound-state problems using the SDP algorithm described previously. 

To be precise, let $V(x) = x^2$ for $x \geq 0$ and zero otherwise. Consider the half line $x \geq 0$ where the Schrodinger equation is
\begin{equation}
    -\psi''(x) + x^2\psi(x) = E\psi(x)
\end{equation}
in units where $\hbar = 2m = 1,\ \omega = 2$. The general solution is given by parabolic cylinder functions $U(a,z)$ (see chapter 12 in \cite{NIST:DLMF}) as
\begin{equation}
    \psi(x) = c_1 U(-E/2, \sqrt{2}x) + c_2U(E/2,i\sqrt{2}x)
\end{equation}
Asymptotically, one finds that only the solution with real argument is normalizable as $x \to \infty$. Up to a formal normalization constant, the feasible solution is
\begin{equation}
    \psi(x) \sim U(-E/2,\sqrt{2}x)
\end{equation}
The values of the function $U$ and its derivative at zero are known; to construct the Robin function $a(E)$ we take their ratio. This no longer depends on the normalization and is given by
\begin{equation}
    \label{eq:SHOafunc}
    a(E) = -\frac{\psi(0)}{\psi'(0)} = \frac{1}{2}\frac{\Gamma\left(\frac14 - \frac{E}{4}\right)}{\Gamma\left(\frac34 - \frac{E}{4}\right)}
\end{equation}
This function already contains all the spectral information of the harmonic oscillator. The full harmonic potential on $\RR$ is even, so all states are of odd or even parity. These states satisfy Dirichlet (D) or Neumann (N) conditions at the origin $x = 0$ respectively. Therefore, the odd states of the harmonic oscillator should correspond to $a = 0$ and the even states to $a = \infty$. 

The Robin function \eqref{eq:SHOafunc} is zero precisely when the gamma function in the denominator diverges and infinite when the numerator diverges. This leads to two quantization rules for the energies $E_n$ depending on the boundary condition in question:
\begin{equation}
    E_n = \begin{cases}
        3 + 4n; & (\operatorname{D}),\\
        1 + 4n; & (\operatorname{N}),
    \end{cases}
\end{equation}
for $n\geq 0$ in each. Together these rules describe the entire harmonic spectrum, which is at $E_n = 2n + 1$ for $n\geq 0$ in our units. 

\begin{figure}[h]
    \centering
    \includegraphics[width = \columnwidth]{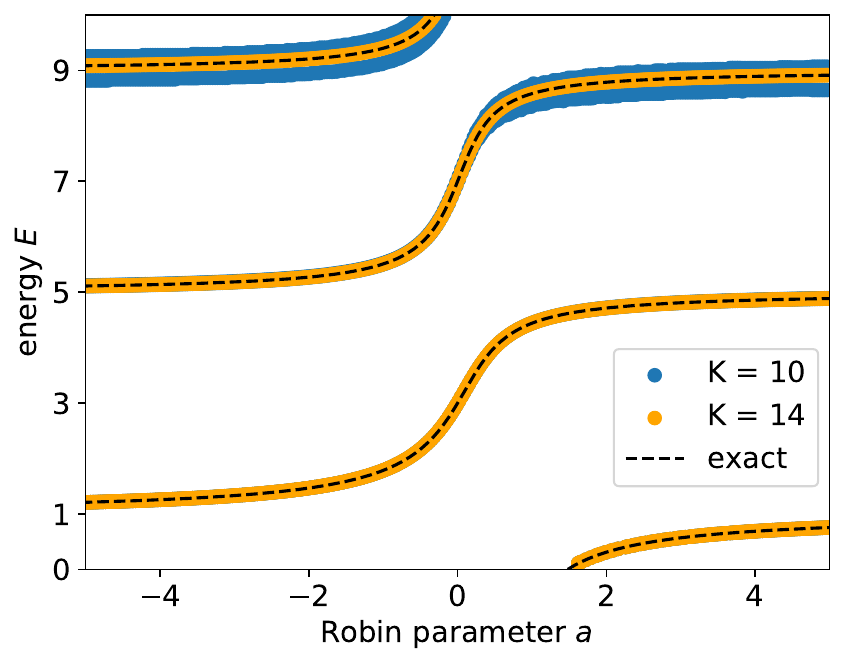}
    \caption{Bootstrapped harmonic spectrum versus Robin parameter for two depths $K = 10,\ 14$. At higher energies the intervals of positivity are larger; their size shrinks as $K$ increases. The exact relation \eqref{eq:SHOafunc} is shown in black. }
    \label{fig:sho_spec}
\end{figure}
We may compare the function $a(E)$ determined here to the bootstrap results. Using the SDP algorithm described in the previous section, we determine the first few harmonic energy levels $E_n(a) \in [0,10]$ for a range of Robin parameter values $a \in [-5,5]$. We then invert the spectrum $E_n(a)$ to find the (single-valued) function $a(E)$. These data and the exact curve are shown in Fig. \ref{fig:sho_spec}, showing excellent agreement.

\subsection{Resonant scattering}
The function $a(E)$ contains information not just about bound states but also metastable ones. Consider the Hamiltonian 
\begin{equation}
    \label{eq:pot_metastable}
    H = p^2 + \frac{1}{2}x^2(x-3)^2
\end{equation}
on the half line $x \geq 0$. This potential has a (local) minimum at $x = 3$ where $V = 0$. Such a local minimum will tend to confine states for a short amount of time, with tunneling to $x < 0$ leading to metastability. 

It is generally known that bound and metastable states are reflected in the analytic structure of the $S$-matrix or, in our case, the reflection coefficient \cite{sakurai_napolitano_2017}. In the complex $E$ plane, bound states correspond to purely imaginary poles in the upper half-plane and metastable states to poles in the lower half plane. 

The signatures of these metastable poles are reflected in the phase shift $\delta(E)$, and in particular its derivative. Let such a metastable pole occur at $E_m = E_0 - i\Gamma$ where $\Gamma > 0$ is the \textit{width} of the metastable state; its characteristic lifetime in the time-dependent picture. Near $E \approx E_m$, the reflection coefficient $R(E)$ should reflect this singular structure and be a pure phase. On general grounds, this implies that near $E_m$ one has
\begin{equation}
    R(E) \approx \frac{E - E_0 - i\Gamma}{E - E_0 + i\Gamma}.
\end{equation}
The phase shift is $\delta = \arg R$; with the identity \eqref{eq:identity} one finds that the derivative of $\delta(E)$ near the resonant energy $E_m = E_0 - i\Gamma$ behaves as
\begin{equation}
    \delta'(E) \approx \frac{2\Gamma}{(E-E_0)^2 + \Gamma^2},
\end{equation}
forming a Lorentz/Cauchy distribution centered at $E_0$ with width $\Gamma$. In the context of decay widths, this is often referred to as the Breit-Wigner distribution, and is studied in numerous texts \cite{landau2013quantum,sakurai_napolitano_2017}. 

\begin{figure}[h]
    \centering
    \includegraphics[width = \columnwidth]{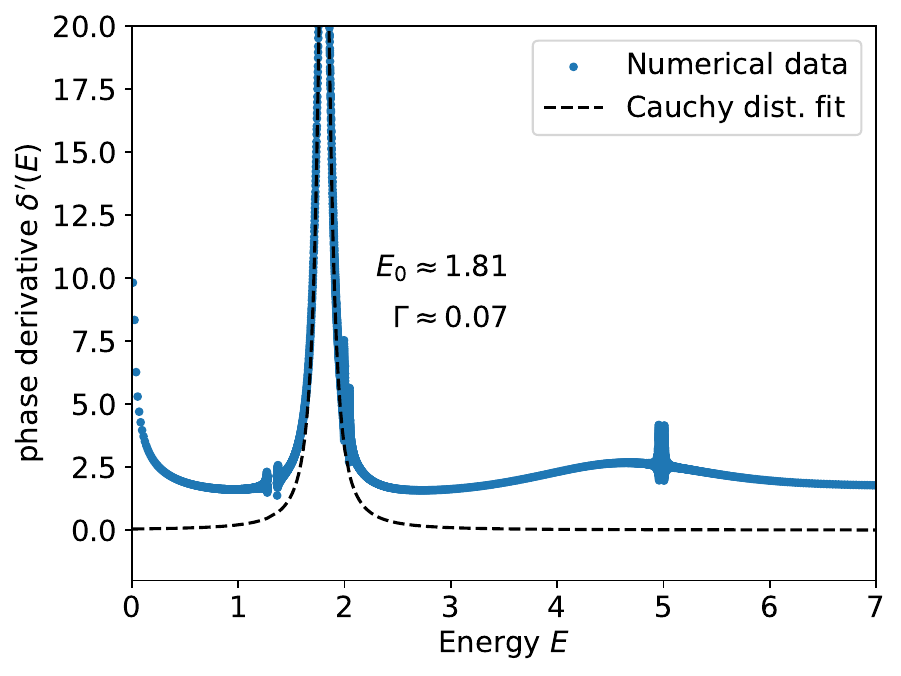}
    \caption{A strong energy resonance in the phase shift for a potential with metastable states. A regression to a Cauchy distribution is included in black. In blue, the transformation of spectral numerical data.}
    \label{fig:cauchy}
\end{figure}
As before, the phase shift may be numerically determined by the spectral flow of $H$ for various values of the Robin parameter $a$, with the results shown in Fig. \ref{fig:cauchy}. Here, we do the numerics using the \textit{Mathematica} function \verb|NDEigensystem| to solve for the lowest three eigenstates of the potential \eqref{eq:pot_metastable} at a range of Robin parameter values $a \in [-20,-20]$. We invert the spectrum to obtain the function $a(E)$, then construct the phase shift using Eqn. \eqref{eq:generalPhase} and its derivative numerically, using a spline interpolation to decrease noise. An energy resonance is clear in Fig. \ref{fig:cauchy}, corresponding to the existence of a metastable state of the potential and consistent with physical expectations. 

\section{Liouville scattering}
The potentials considered in the previous sections are unique and hand-picked in that they are identically zero on the negative real axis. This allows an exact solution of the Schrodinger equation in the $x<0$ region  and subsequent matching of boundary conditions to extract the phase shift from dependence of the Robin parameter on the energy. 

In many situations, the potential may be nonzero but slowly-varying and weak at large negative $x$. This is the case for the exponential potential which famously arises as the potential in the semiclassical Liouville theory \cite{NAKAYAMA_2004,Seiberg:1990eb}. To be precise, after canonical quantization the zero-mode Hamiltonian is
\begin{equation} 
    \label{eq:LiouvilleHam}
    H_0 = \frac{p^2}{2} + 2\pi\mu e^{2bx},
\end{equation}
ignoring any zero-point energy. From the perspective of the fundamental dilaton field this Hamiltonian follows by considering spatially uniform solutions: the `mini-superspace' approximation. We are not concerned with the physical origin of the theory so much as applying the tools of the previous section to the one-sided exponential potential. 

The Schrodinger equation corresponding to the Hamiltonian \eqref{eq:LiouvilleHam} is
\begin{equation}
   -\frac12 \psi''(x) + 2\pi \mu e^{2bx}\psi(x) = \frac{k^2}{2}\psi(x),
\end{equation}
where now we use the dispersion $E = k^2/2$. This is solved by a linear combination of Bessel functions whose properties are well-known. By demanding regularity at $x = \infty$ and matching to the plane wave ansatz \eqref{eq:planewave}, one find that the reflection coefficient $R(k)$ is given by \begin{equation}
    \label{eq:louexactphase}
    R(k) = -(\pi \mu/b^2)^{-ik/b}\frac{\Gamma(1+ik/b)}{\Gamma(1-ik/b)}
\end{equation}
We will reproduce this, at least approximately, by combining the bound-state approach with a WKB approximation to the wavefunction. One sets a boundary and solves a one-sided bound state problem with Robin boundary condition on the right-hand side of the boundary, as before. Placing the boundary at some modestly large, negative position $L$ then ensures that the semiclassical (WKB) approximation is valid and accurate to the left of the boundary, at least for potentials which smoothly go to zero. This is certainly true for the exponential potential $V = e^{2bx}$, which, along with its derivative, goes to zero very quickly as $x \to -\infty$. For our purposes, we need a regime where $k^2\gg V(x)$ and that $V$ varies slowly on the scale of $k$ (any feature of $V$ needs to occur on a region that is long compared to local wavelength $1/k(x)\sim 1/k$). 

\subsubsection*{The WKB wavefunction}
Any wavefunction can be written in the form
\begin{equation}
    \label{eq:WKBansatz}
    \psi(x) \propto \exp(\lambda(x)+i\omega(x)).
\end{equation}
In the WKB approximation, the functions $\lambda(x),\ \omega(x)$ (modulating the amplitude and phase, respectively) are expanded in powers of the formally small parameter $\hbar$ as
\begin{align}
    \omega(x) &= \frac{1}{\hbar}S_0(x) + \hbar S_2(x) + \OO(\hbar^3),\\
    \lambda(x) &= S_1(x) + \hbar^2 S_3(x) + \OO(\hbar^4).
\end{align}
The functions $S_n(x)$ satisfy a hierarchy of equations found by substituting the ansatz \eqref{eq:WKBansatz} into the Schrodinger equation and solving order-by-order in $\hbar$. In particular, the lowest-order correction $S_0(x)$ solves the eikonal equation
\begin{equation}
    [S_0'(x)]^2 = 2(E-V(x)).
\end{equation}
If one takes $V \to 0$ in this equation, one finds $S_0(x) = \pm \sqrt{2E}x = \pm kx$, reproducing the free plane-wave solution seen earlier. 

Because the potential is real, on general grounds the semiclassical (WKB) wavefunction is given by the linear combination
\begin{equation}
    \label{eq:lambdaansatz}
    \psi_{scl} = Ae^{\lambda(x)}e^{i\omega(x)} + Be^{\lambda(x)}e^{-i\omega(x)}.
\end{equation}
The data of the reflection coefficient $R(k)$ is contained in the asymptotic relative normalization of these two solutions: roughly the ratio $B/A$ as $x \to -\infty$. The semiclassical wavefunction interpolates between the asymptotic limit, where the $S$-matrix is defined, and a finite value $L < 0$, where the potential is slowly varying. Here where the WKB approximation is valid we place a boundary at $x = L < 0$. On the right-hand side, one defines a bound state problem with Robin boundary condition
\begin{equation}
    \psi(L) + a_L \psi'(L) = 0.
\end{equation}
Solving this determines a spectrum $E_n(a_L)$ which is subsequently inverted to obtain a function $a_L(E)$ as described in the previous section.

By continuity of the wavefunction and its derivative at the point $x = L$, the semiclassical wavefunction should define the same Robin parameter:
\begin{equation}
    a_L = -\frac{\psi_{scl}(L)}{\psi'_{scl}(L)}.
\end{equation}
Substituting the equation \eqref{eq:lambdaansatz}, one can determine a relation between the ratio of coefficients $B/A$ of the semiclassical wavefunction and the Robin parameter $a_L$ at finite $L$:
\begin{equation}
    B/A = -e^{2i\omega(L)}\frac{1+a_L\lambda'(L) + ia_L\omega'(L)}{1+a_L\lambda'(L) - ia_L\omega'(L)}
\end{equation}
where the $a_L,\ \lambda(x)$, and $\omega(x)$ all implicitly depend on the energy $E$. 

On the other side of the domain, relating the reflection coefficient $R$ to the ratio $B/A$ requires another matching formula. Define the angle $\theta_0$ as the limit
\begin{equation}
    \theta_0 \equiv \lim_{x\to-\infty}\omega(x) - kx
\end{equation}
recalling that here $k = \sqrt{2E}$. This limit is finite in general due to the known asymptotic behavior of the function $S_0(x)$, which dominates the expansion of the function $\omega(x)$ at large negative $x$. With this angle appearing as a decorating phase, the asymptotic reflection coefficient is
\begin{equation}
    R_{scl}(k) = -e^{-2i\theta_0}e^{2i\omega(L)}\frac{1+a_L\lambda'(L) + ia_L\omega'(L)}{1+a_L\lambda'(L) - ia_L\omega'(L)}
\end{equation}
Taking the argument of this function yields the WKB-corrected phase shift as a function of the wavenumber $k = \sqrt{2E}$:
\begin{multline}
    \label{eq:semiclass_phase}
    \delta_{scl}(k) = -\pi  + 2\tan^{-1}\left(\frac{a_L \omega'(L)}{1+a_L \lambda'(L)}\right)\\ + + 2(\omega(L) - \theta_0) \mod 2\pi
\end{multline}
The WKB approximation may then be carried out to any order in calculating the functions $\omega(x)$ and $\lambda(x)$ to determine increasingly accurate expressions for the phase shift. 

As the potential $V \to 0$, one finds that $\lambda' \to 0$ while $\omega' \to k$. In this limit one also has $\omega(L) - \theta_0 \to kL$. This limit reproduces the expression \eqref{eq:generalPhase} for a boundary at $x \neq 0$. In this way the zero-order WKB approximation reduces to the results determined earlier. 

\subsubsection*{The exponential potential}
As a demonstration, we show the accuracy of the WKB approach to approximating the scattering phase \eqref{eq:louexactphase}, setting $\mu = 1/2\pi$ and $b = 1$ for definiteness. First, one can explicitly compute the lowest-order WKB functions $S_n(x)$, from which the amplitude and phase modulation may be constructed. Defining the semiclassical momentum
\begin{equation}
    p(x) = \sqrt{2(E - V(x))} = \sqrt{2(E-e^{2x})},
\end{equation}
the first four WKB functions $S_n(x)$ are given by:
\begin{align}
    S_0(x) &= p(x) -k\tanh^{-1}\left(\frac{p(x)}{k}\right),\\
    S_1(x) &= -\frac12 \log p(x),\\
    S_2(x) &= \frac{2E + 3e^{2x}}{24\sqrt{2}p(x)^3},\\
    S_3(x) &= \frac{p''(x)}{8p(x)^3} - \frac{3p'(x)^2}{16p(x)^4}.
\end{align}
While the corrections $S_1,\ S_3$ are universal, the corrections $S_0,\ S_2$ depend on the potential. The correction $S_0$ is by far the largest in absolute numerical terms and increases in size as $x \to -\infty$. A large phase is to be expected, as the eikonal or first-order WKB approximation is equivalent to a saddle-point or stationary phase path integral. 

\begin{figure}
    \centering
    \includegraphics[width = \columnwidth]{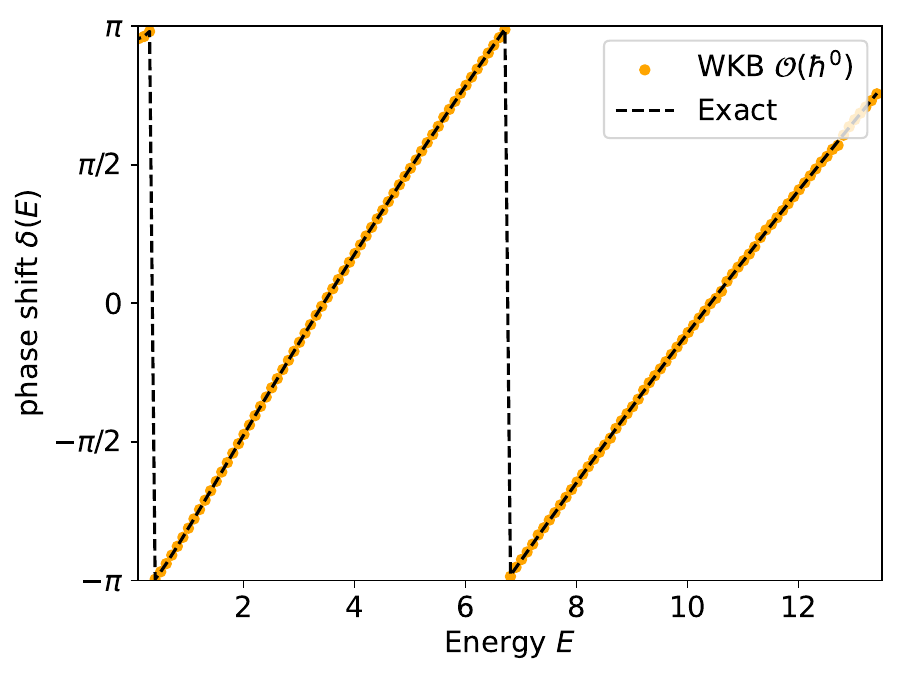}
    \caption{Phase shift due to reflection off the exponential potential reconstructed with the first-order WKB approximation (orange) compared to the exact result (dashed, black). With a boundary at $L = -5$, we solve the bound state problem on the right to define the Robin function $a_L(E)$. In blue, the naive application of Eqn.~\eqref{eq:generalPhase} with $a_L(E)$. In orange,
    the correction functions $S_0(x),\ S_1(x)$ give the approximate semiclassical phase shift with good accuracy.}
    \label{fig:wkb}
\end{figure}

Most applications of the WKB method include only the first two corrections $S_0,\ S_1$. Indeed we find that these corrections are more than sufficient to reproduce the exact phase shift law \eqref{eq:louexactphase} by solving bound-state problems and applying the approximation of Eqn. \eqref{eq:semiclass_phase}. These data are shown in Fig. \ref{fig:wkb} along with the exact curve for a boundary at $L = -5$. 

\section{Conclusion}
In this letter we have described how to apply numerical solutions to bound-state problems, with Robin boundary conditions, to the numerical solution of scattering off one-sided potentials. The intuitive picture is that just as bound states are equivalent to properly interfering sets of scattering states, the information of scattering states can be represented as continuous families of bound states. We gave examples of this correspondence, comparing numerical and analytical results.  We also described a semiclassical generalization using the WKB approximation. 

In the context of previous work on the quantum-mechanical bootstrap, this letter adds control over the problem of scattering, thereby almost completing the bootstrap's application to `textbook' one-dimensional quantum mechanics. We say \textit{almost} as there are problems naturally suggested here that we have not addressed. Most notably these include computing not only reflection but also transmission amplitudes, and from these constructing a non-trivial $S$-matrix. From the perspective of this work, determining transmission and reflection would involve solving a bound-state problem on a finite interval. On such a domain---the traditional home of Sturm-Liouville theory---anomalies arise at both boundaries dependent on the (independent) conditions ascribed to each boundary. In principle, solving bound state problems on the finite interval is well understood, and is easily approached by finite-element methods or by spectral methods. In order to complete the discussion of scattering one should be able to characterize both reflection and transmission entirely in terms of a bound state problem in an interval in analogy to the present discussion of pure reflection. This generalization is beyond the scope of this letter.

{\em Acknowledgements:} 
D.B. Research supported in part by the Department of Energy under Award No. DE-SC0019139

\bibliographystyle{apsrev4-1}
\bibliography{refs}

\end{document}